\documentclass[12pt]{article}
\usepackage{graphicx,ccaption}
\def\simless{\mathbin{\lower 1pt\hbox
   {$\spose{\raise 5pt\hbox{$\char'074$}}\char'430$}}}
\def\simgreat{\mathbin{\lower 1pt\h   {$\spose{\raise 5pt\hbox{$\char'076$}}\char'430$}}}
\def\simgreat{\gapp}
\def\simless{\lapp}
\def\lapp{\mathbin{\raise2pt \hbox{$<$} \hskip-9pt \lower4pt \hbox{$\sim$}}}
\def\gapp{\mathbin{\raise2pt \hbox{$>$} \hskip-9pt \lower4pt \hbox{$\sim$}}}
\newcommand{\rfnce}{\par \noindent \hangindent 15pt{}}

\newcommand{\vv}{\mbox{\boldmath $v$}}
\newcommand{\nhat}{\hat{n}}
\newcommand{\vnhat}{\hat{\mbox{\boldmath $n$}}}

\newcommand{\vgrad}{\mbox{\boldmath $\nabla$}}
\newcommand{\vclp}{\mbox{\boldmath $\mathcal{P}$}}
\newcommand{\ud}{\mathrm{d}}

\captionnamefont{\footnotesize\it}
\captiontitlefont{\footnotesize}
\captiondelim{. }

\begin{document}

\title
{\large \bf CONSTRAINING FUNDAMENTAL PHYSICS WITH THE COSMIC MICROWAVE BACKGROUND\footnote{ Presented at the {\it Workshop on Cosmology and Gravitational
Physics}, 15--16 December 2005,
 Thessaloniki, Hellas (Greece), {\it Editor}: N.\ K.\ Spyrou.}}
 \author {\small\bf Anthony Challinor\thanks{a.d.challinor@mrao.cam.ac.uk}\\
\small $^1$ Astrophysics Group, Cavendish Laboratory, J.\ J.\ Thomson
Avenue,\\
\small Cambridge, CB3 0HE, UK}
\date{}
\maketitle


\begin{abstract} 

The temperature anisotropies and polarization of the cosmic microwave
background (CMB) radiation provide a window back to the physics of the early
universe. They encode the nature of the initial fluctuations
and so can reveal much about the physical mechanism that led to
their generation. In this contribution we review
what we have learnt so far
about early-universe physics from CMB observations, and what we hope to learn
with a new generation of high-sensitivity, polarization-capable instruments.

\end{abstract}

\vspace{1cm}
\noindent{\bf {1. Introduction}}\\

The cosmic microwave background (CMB) radiation has an almost perfect
blackbody spectrum with mean temperature 2.725~K~(Mather et al 1994).
Once corrected for
a kinematic dipole due to our motion, the CMB is remarkably isotropic
with fluctuations at the $10^{-5}$ level. These were first detected by the
COBE satellite in 1992~(Smoot et al 1992)
and have now been measured
over three
decades of scale by a combination of ground-based and balloon-borne
experiments and the WMAP
satellite~(Bennett et al 2003). CMB photons were tightly coupled
to matter in the early universe but propagated essentially freely after
the primordial plasma recombined to form neutral atoms\footnote{Recent
WMAP3 results indicate that around 10\%
of the photons re-scattered once the universe reionized around redshift 10.}.
For this reason, the CMB provides a snapshot of the spatial fluctuations
on a spherical shell of comoving radius 14000~Mpc at redshift $z\sim 1000$
when the universe was only $400\times 10^3$ years old. The small amplitude
of these fluctuations [$O(10^{-5})$] means they are well described by linear
perturbation theory and the physics of the CMB is thus very well
understood.

The fluctuations on the last-scattering surface are believed to have
resulted from primordial curvature fluctuations
plausibly generated quantum mechanically during an inflationary phase
in the first $10^{-35}$~seconds. These primordial fluctuations are processed
by gravitational instability and, on smaller scales, by the acoustic
physics of the plasma which is supported by photon pressure. This makes the
CMB anisotropy sensitive to the physics that initially produced the
fluctuations and to the composition of the matter in the universe which
affects the acoustic physics.

In this \emph{contribution} we review what we have learnt from current
CMB observations, with particular emphasis placed on constraints on
fundamental physics and models of the early universe. We also look forward
briefly to what might be learnt with more sensitive future observations,
such as tighter constraints on the gravitational wave background predicted
from inflation and sub-eV constraints on neutrino masses from weak
gravitational lensing of the CMB.

\vspace{1cm}
\noindent{\bf {2. CMB Physics}}\\

We begin with a brief review of the physics of the CMB temperature
and polarization anisotropies. For more complete reviews see e.g.\
Hu \&\ Dodelson (2001), Hu (2002) and Challinor (2005).

Consider a spatially-flat background universe linearly perturbed
by density perturbations and gravitational waves. The metric can be taken to
be $\ud s^2 = a^2(\eta)\{(1+2\psi) \ud\eta^2-[(1-2\phi)\delta_{ij} + h_{ij}]
\ud x^i \ud x^j\}$, where the Newtonian-like potentials $\psi$ and $\phi$
describe the scalar (density) perturbations and the
transverse, trace-free $h_{ij}$
describes tensor perturbations (gravitational waves). In the absence of
anisotropic stresses $\phi=\psi$. If we ignore scattering after the universe
reionized, and approximate the last-scattering surface as sharp,
the fractional temperature fluctuation along a line of sight $\vnhat$ has
a scalar contribution
\begin{equation}
\Theta(\vnhat) = \Theta_0 |_E + \psi |_E - \vnhat\cdot \vv_b |_E +
\int_E^R(\dot{\psi}+\dot{\phi})\,\ud\eta \, ,
\label{eq:1}
\end{equation}
and a tensor contribution
\begin{equation}
\Theta(\vnhat)=-\frac{1}{2} \int_E^R \dot{h}_{ij} \nhat^i \nhat^j \, \ud\eta
\, .
\label{eq:2}
\end{equation}
Here, an overdot denotes a derivative with respect to conformal time
$\eta$.
Performing
a spherical harmonic expansion of the anisotropies,
\begin{equation}
\Theta(\vnhat) = \sum_{lm} a_{lm} Y_{lm}(\vnhat) \, ,
\label{eq:3}
\end{equation}
the power spectrum is defined by $\langle a_{lm} a^*_{l'm'}\rangle
= C_l^T \delta_{ll'} \delta_{mm'}$ where the angle brackets denote
an ensemble average (or quantum expectation value). This form of
the correlator follows from statistical isotropy and
homogeneity. The quantity $l(l+1)C_l^T/2\pi$ is usually plotted and this
gives the contribution to the mean-square anisotropy per $\ln l$.

\begin{figure}
\begin{center}
\includegraphics[width=0.45\textwidth,angle=-90]{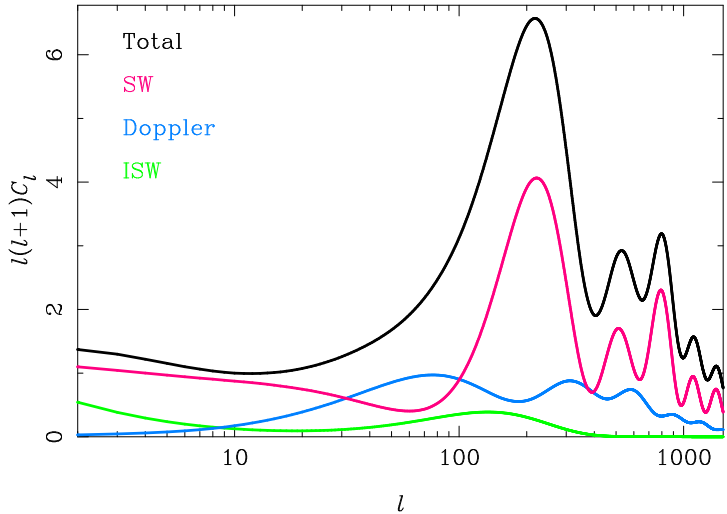}
\end{center}
\caption{\label{challinor:fig1} Scalar contributions to the temperature
power spectrum from scale-invariant adiabatic initial fluctuations. At high
$l$ the contributions are (from top to bottom): total power (black);
Sachs-Wolfe (magenta); Doppler (blue);
and the integrated Sachs-Wolfe effect (green).}
\end{figure}

The scalar anisotropies, equation~(\ref{eq:1}),
are sourced by fluctuations on the last-scattering
surface at position $E$ and an integral along the line of sight
involving the evolution of the Weyl potential $\psi+\phi$. Regions
of photon over-density have a positive intrinsic
temperature fluctuation $\Theta_0$, but this is modified by the
local gravitational potential $\psi|_E$ that the photon must climb out of
at last scattering to give the observed anisotropy. There is a further
contribution from the baryon peculiar velocity $\vv_b$ at the
last-scattering event which Doppler blueshifts photons that scatter
along the direction of motion. The last `integrated Sachs-Wolfe' (ISW) term
in equation~(\ref{eq:1}) contributes to the large-angle fluctuations
because of the decay of the gravitational potentials once dark energy
dominates the expansion, and also on smaller scales due to evolution
of the potentials around last scattering before the universe
is fully matter dominated. The various contributions to the anisotropy
power spectrum from scalar perturbations are shown in
Fig.~\ref{challinor:fig1} for \emph{adiabatic initial conditions}
where the \emph{relative} composition of the universe is initially
the same everywhere. Such initial conditions are natural in inflation
models with only a single field, but in models with multiple fields
the initial conditions may include an \emph{isocurvature} component
where the relative composition fluctuates in space in such a way as
not to perturb the curvature.

The initial curvature fluctuations are processed by gravitational and acoustic
physics to determine the fluctuations on the last scattering surface.
In particular, on scales above the photon mean free path, the radiation
and baryons are tightly-coupled and the radiation is isotropic in the
plasma rest-frame. In this limit, the intrinsic temperature fluctuation
satisfies an oscillator equation~(Hu \&\ Sugiyama 1995)
\begin{equation}
\ddot{\Theta}_0 + \frac{\mathcal{H}R}{1+R}\dot{\Theta}_0 -
\frac{1}{3(1+R)}\vgrad^2 \Theta_0 = 4\ddot{\phi}+ \frac{4\mathcal{H}R}{1+R}
\dot{\phi} + \frac{4}{3} \vgrad^2 \psi \, ,
\label{eq:4}
\end{equation}
where $\mathcal{H}\equiv \dot{a}/a$ is the conformal Hubble parameter
and $R\equiv 3\rho_b/(4\rho_\gamma)$ is the ratio of baryon to photon
density. The sound speed in the plasma is $c_s^2 = 1/[3(1+R)]$ and is reduced
from the value for a photon gas by the inertia of baryons.
The oscillator is damped by the Hubble drag on the baryons\footnote{%
Expansion redshifts away the peculiar velocity of matter.} and is driven
by the gravitational potentials. A potential well accretes surrounding
plasma until the induced pressure gradients balance the gravitational
force. At this point $\Theta_0 = -(1+R)\psi$ (ignoring evolution of the
gravitational potentials) but the plasma is still infalling and doesn't
turn around until a higher density. The amplitude of the oscillation about
the midpoint is determined by the initial conditions and gravitational
driving; for adiabatic initial conditions the plasma starts off over-dense in
potential wells with $\Theta_0 = -\psi(0)/2$ \emph{for all wavelengths}.
This means that the oscillations of different Fourier modes all start off in
phase, at an extrema, but the period of their subsequent oscillation is
wavelength dependent. It follows that at the last-scattering surface
different wavelengths will be caught at different phases of a cosine-like
oscillation.
Those wavelengths at an extremum at last scattering will give rise to a
peak in the anisotropy power spectrum at a multipole $l \sim k d_A(r_*)$,
where $d_A(r_*)$ is the angular-diameter distance back to last
scattering at comoving distance $r_* \approx 14\,\mathrm{Gpc}$. 

The tensor anisotropies, equation~(\ref{eq:2}), involve an integral of the
metric shear $\dot{h}_{ij}$
which describes the local anisotropy in the expansion of the universe due
to the presence of gravitational waves: CMB photons experience
line-of-sight-dependent redshifts that imprint temperature anisotropies
on the microwave sky. A cosmological background of gravitational waves
is naturally produced during inflation from the vacuum fluctuations in
the massless field $h_{ij}$~(Starobinskii 1979);
see Sec.~5. On scales larger
than the Hubble radius, $h_{ij}$ is constant but once a mode enters
the Hubble radius it starts to oscillate with an amplitude that
decays as $a^{-1}$. On scales $\simless 3^\circ$, i.e.\
$l \simgreat 60$,
the relevant gravitational waves
at any distance back to last scattering are sub-Hubble and so the tensor
signal is only appreciable on scales larger than this.

\vspace{1cm}
\noindent{\bf {2.1 CMB Polarization}}\\

The other important CMB observable is its linear polarization~(Rees 1968).
The tightly-coupled description of the plasma given above breaks down
for short wavelength modes, and universally as the mean-free path grows
around recombination. As photons start to diffuse out of over-dense regions
their density and velocity perturbations are damped which leads to
an exponential damping tail in the anisotropy power spectrum~(Silk 1968).
However, a quadrupole anisotropy also starts to develop in the radiation
and subsequent Thomson scattering generates linear polarization with an r.m.s.\
$\sim 5\,\mu\mathrm{K}$.

\begin{figure}
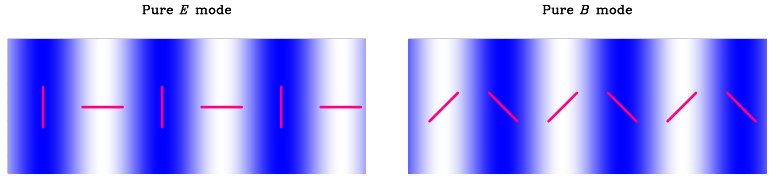

\begin{center}
\includegraphics[width=0.15\textwidth,angle=-90]{planeEmode.ps}
\quad
\includegraphics[width=0.15\textwidth,angle=-90]{planeBmode.ps}
\end{center}
\caption{\label{challinor:fig2} Polarization patterns for a pure
$E$ mode (left) and $B$ mode (right) for a single Fourier mode
on a flat patch of sky. In the basis defined by the wavevector, the
$E$ mode has vanishing $U$ and the $B$ mode vanishing $Q$.}
\end{figure}

Linear polarization is described by Stokes parameters $Q$ and $U$ which measure
the difference in intensity between two orthogonal polarizers ($Q$) and the
same rotated by $45^\circ$. The Stokes parameters form the components
of the polarization tensor $\vclp$ which measures the (zero-lag) correlation
of the
electric field components in the radiation. They are 
therefore basis dependent but coordinate-independent fields can be
derived by writing $\vclp$ as a gradient (or electric) part and curl (or
magnetic) part (Kamionkowski, Kosowsky \& Stebbins 1996; Zaldarriaga
\& Seljak 1996):
\begin{equation}
\mathcal{P}_{ab}(\vnhat) = \sum_{lm} \sqrt{\frac{(l-2)!}{(l+2)!}}
\left( E_{lm} \nabla_{\langle a} \nabla_{b \rangle} Y_{lm}(\vnhat)
+ B_{lm} \epsilon^c{}_{\langle a}\nabla_{b\rangle}\nabla_c Y_{lm}(\vnhat)
\right)
\, .
\label{eq:5}
\end{equation}
Here, the covariant derivatives are in the surface of the unit (celestial)
sphere, angle brackets denote the symmetric, trace-free part and we
have expanded the electric and magnetic parts in spherical harmonics.
By way of example, Fig.~\ref{challinor:fig2} shows the polarization
patterns for $E$ and $B$ modes that are locally plane waves.
The $E$ field behaves as a scalar under parity transformations but
$B$ is a pseudo-scalar.
If parity-invariance is respected in the mean, there are only three
additional non-vanishing power spectra: the $E$- and $B$-mode
auto-correlations, $C_l^E$ and $C_l^B$, and the cross-correlation of
$E$ with the temperature anisotropies $C_l^{TE}$. For linear scalar
perturbations the $B$-mode polarization vanishes by symmetry, but
gravitational waves produce $E$ and $B$ modes with approximately equal
power~(Kamionkowski et al 1996; Zaldarriaga \& Seljak 1996; Hu \& White 1997).
This makes the large-angle $B$ mode of polarization an
excellent
probe of primordial gravitational waves and several groups are now
developing instruments with the aim of searching for this signal; see
Sec.~6.

\begin{figure}
\begin{center}
\includegraphics[width=0.45\textwidth,angle=-90]{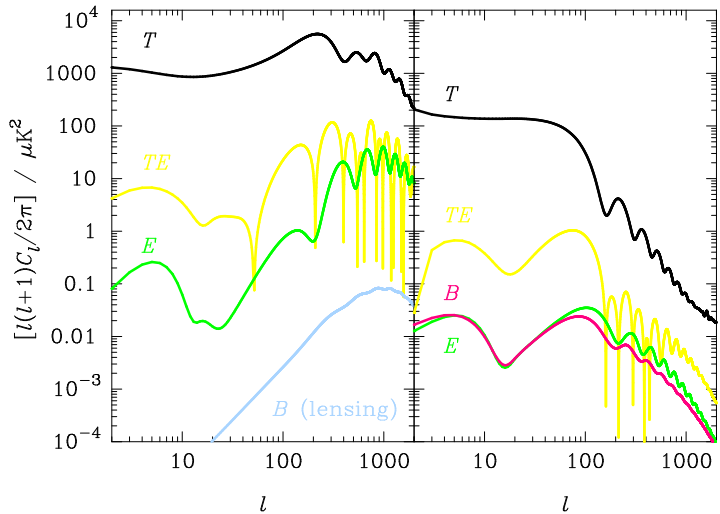}
\end{center}
\caption{\label{challinor:fig3} CMB temperature and polarization
power spectra from scalar
(left) and tensor perturbations (right) for a tensor-to-scalar ratio
$r=0.38$. The $B$-mode power generated by weak gravitational
lensing is also shown.}
\end{figure}

The temperature and polarization power spectra from density perturbations and
gravitational waves are compared in Fig.~\ref{challinor:fig3}. 
The gravitational wave amplitude is set to a value close to the current upper
limit from the temperature anisotropies. Note that for scalar perturbations,
$C_l^E$ peaks at the troughs of $C_l^T$ since the radiation quadrupole
at last scattering is derived mostly from the plasma bulk velocity which
oscillates $\pi/2$ out of phase with the intrinsic temperature. The
polarization from density perturbations is maximised around $l \sim 1000$
which is
related to the angle subtended by the photon mean-free patch around
recombination. The `bump' in the polarization spectra on large angles
is due to reionization~(Zaldarriaga 1997).
Once the ultraviolet light from the first compact objects has
reionized the intergalactic medium, the liberated electrons can re-scatter
the CMB. This has the effect of damping the temperature and polarization
spectra by $e^{-2\tau}$, where $\tau$ is the optical depth to Thomson
scattering, on scales
inside the horizon at that epoch. However, Thomson scattering of the
quadrupole anisotropy that has now developed in the free-streaming radiation
produces new large-angle polarization. The power in the reionization bump
scales like $\tau^2$ and the angular scale is related to the epoch of
reionization. The high value $\tau=0.17$ adopted in
Fig.~\ref{challinor:fig3}, favoured by the first-year WMAP data
(Kogut et al 2003), is now at odds with the recent three-year WMAP data (Page
et al 2006); see Sec.~4.
Also shown in the figure is the non-linear $B$-mode signal generated
by weak gravitational lensing of the primary $E$-mode polarization from
density perturbations~(Zaldarriaga \& Seljak 1998; Lewis \& Challinor 2006).
We discuss this signal further in Sec.~6.

\vspace{1cm}
\noindent{\bf {3. Current Observational Situation}}\\

\begin{figure}
\begin{center}
\includegraphics[width=0.34\textwidth,angle=90]{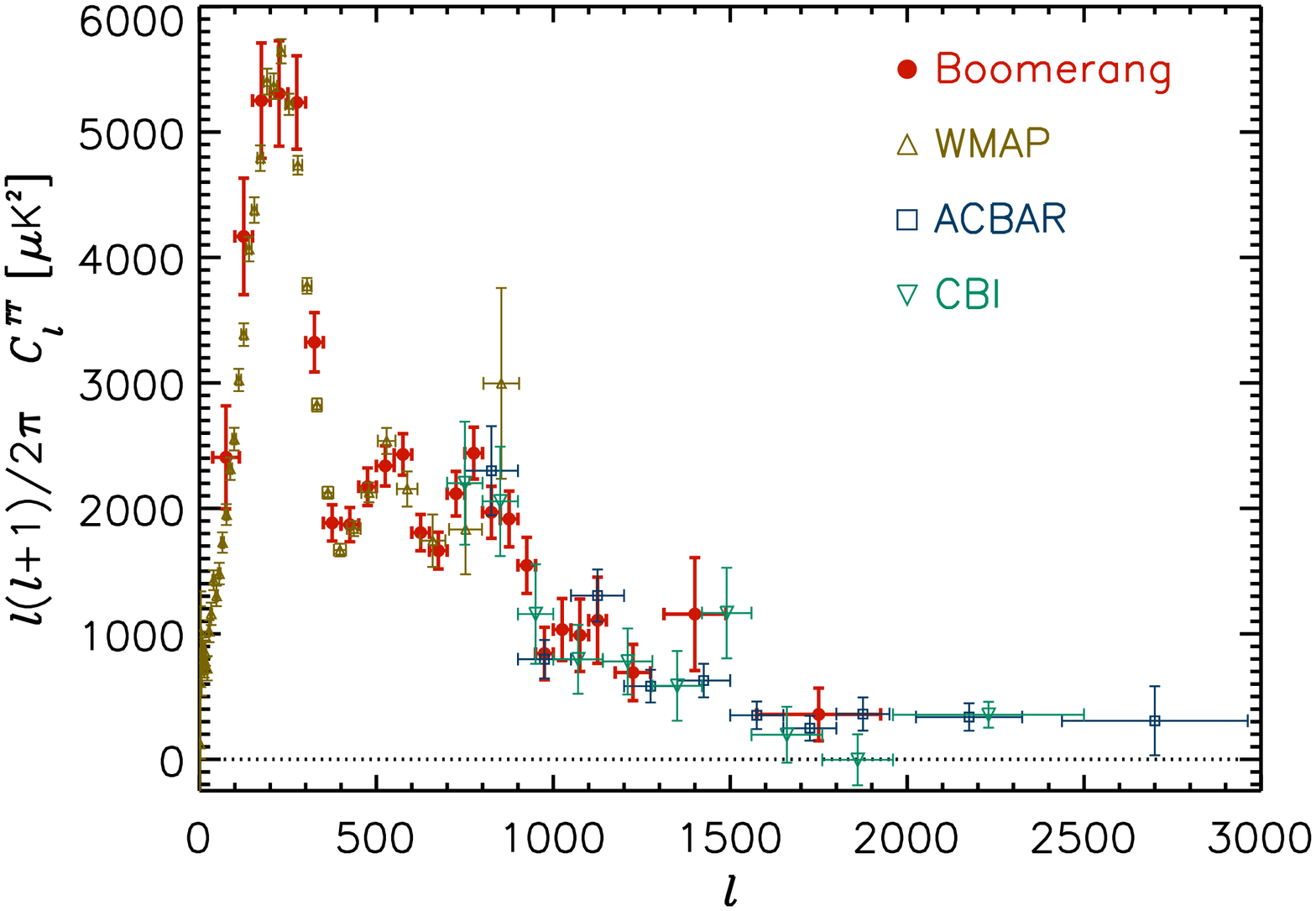}
\raisebox{1.0\totalheight}{%
\includegraphics[width=0.34\textwidth,angle=-90]{current_data_TEandE_0306.ps}
}
\end{center}
\caption{\label{challinor:fig4} Recent CMB temperature (left) and polarization
(right; $C_l^{TE}$ top and $C_l^E$ bottom) power spectra measurements.
For polarization all measurements are plotted including WMAP3.
The solid lines in the polarization
plots are the theoretical expectation on the basis of the temperature
data and an optical depth $\tau=0.08$.
The temperature
plot, from Jones et al (2005), shows results from WMAP1 and a selection of
ground and balloon-borne instruments.
The recent WMAP3 data helps delimit
the third acoustic peak in $C_l^T$ further.}
\end{figure}

Imaging the CMB temperature anisotropies is now a mature field. A number
of complementary technologies have been deployed to map the CMB
from frequencies of a few tens to hundreds of GHz. The remarkable
full-sky maps from the WMAP satellite, most recently on the basis of
three years of data~(Hinshaw et al 2006),
have a resolution up to 15 arcmin and are signal-dominated
up to multipoles of a few hundred. The main cosmological
information in the CMB images is encoded in their power spectra so
in Fig.~\ref{challinor:fig4} we show a selection of recent
measurements of the $C_l^T$ taken from Jones et al (2005).
The qualitative
agreement with the theoretical expectation is striking, with
the measurements clearly delimiting the first three acoustic peaks.
As we discuss later, the agreement stands up to rigorous statistical analysis
and has provided a very powerful means of constraining cosmological parameters
and models. While the community awaits the high-sensitivity, full-sky results
from the Planck satellite, due for launch in 2008, a number of groups
are working to improve on ground-based measurements of the small angular
scales inaccessible to WMAP.

Polarization measurements are less mature, with the first detection reported
by DASI in 2002. Since that first detection, four further groups
have published detections of $E$-mode polarization through its auto-correlation
$C_l^E$. These measurements, also shown in Fig.~\ref{challinor:fig4},
are still very noisy, but the qualitative agreement with the best-fit
model to the temperature spectrum is striking. 
The cross-correlation $C_l^{TE}$ has also been measured by several of
these groups, and with the arrival of the third-year WMAP data the measurements
are now quite precise for $l \simless 200$. As we discuss further in
Sec.~6, several new high-sensitivity polarization-capable experiments
are currently under construction and these have the ambition of
measuring the $B$-mode power spectrum. At present only upper limits
exist for $C_l^B$; see Fig.~\ref{challinor:fig6}. These instruments
will also significantly improve on current measurements of $C_l^E$.

\vspace{1cm}
\noindent{\bf {4. Major CMB Milestones and their Cosmological Implications}}\\

The current CMB data has confirmed a number of bold theoretical predictions,
some of which pre-date the data by over thirty years. In this section
we briefly describe these major milestones and discuss their implications
for constraining the cosmological model.

\vspace{1cm}
\noindent{\bf {4.1 Sachs-Wolfe Plateau and the Late-time ISW Effect}}\\

The large-angle temperature anisotropies are dominated by the Sachs-Wolfe
effect, $\Theta_0 + \psi$, and the ISW effect~(Sachs \& Wolfe 1967).
For adiabatic initial conditions, the combination $\Theta_0 + \psi$
reduces to $\psi/3$ on scales above the sound horizon at last scattering
and so potential wells appear as cold regions. For a nearly scale-invariant
spectrum of primordial curvature perturbations, we should have
$l(l+1)C_l^T \approx \mathrm{const.}$ on such scales. This plateau was first
seen in the COBE data~(Hinshaw et al 1996), and has since been
impressively verified
by WMAP. Departures from scale-invariance imply a slope to the $l(l+1)C_l^T$
spectrum on large scales and this can be used to help constrain the spectral
index $n_s$ of the primordial spectrum; see Sec.~5. In practice, the
large-angle data is not very constraining due to the large cosmic
variance there, i.e.\ the fact we only have access to a sample of $2l+1$
spherical harmonic modes at each $l$ with which to estimate the variance of
their population, $C_l^T$.

The late-time ISW effect, from the decay of the Weyl potential $\phi+\psi$
as dark energy dominates, contributes significantly on the largest angular
scales. It is suppressed on smaller scales due to peak-trough cancellation
in the integral along the line of sight and from the decay of
the potential on sub-horizon scales during radiation domination. The
late-time ISW adds incoherently with the Sachs-Wolfe contribution since
at a give $l$ they probe different linear scales. It is the only way to probe
late-time growth of structure with linear CMB anisotropies (see
Sec.~7 for an example of a non-linear probe), but this is hampered
by large cosmic variance. The late-ISW effect produces a positive
correlation between
large-angle temperature fluctuations and tracers of the gravitational
potential at redshifts $z\simless 1$. This
was first detected by correlating the one-year WMAP data with
the X-ray background and with the projected number density of radio
galaxies~(Boughn \& Crittenden 2004),
and has since been confirmed with several other tracers
of large-scale structure. The correlation is sensitive to both
the energy density in dark energy and any evolution with redshift. The
ISW constraints on the former indicate $\Omega_\Lambda \sim 0.8$, consistent
with other cosmological probes; see e.g.~Cabre et al (2006).
As yet there is no
evidence for evolution but significant improvements can be expected from
tomographic analyses of upcoming deep galaxy surveys.

\vspace{1cm}
\noindent{\bf {4.2 Acoustic Peaks}}\\

The first three acoustic peaks are clearly resolved by the current
temperature-anisotropy power spectrum. Corresponding oscillations
are also apparent in the $TE$ cross-correlation data and, at lower
significance, in the $EE$ power spectrum. For adiabatic models,
the positions of the peaks are at the extrema of a cosine oscillation,
giving $l \approx n \pi d_A(r_*)/r_s$ where $r_s \equiv \int_0^{\eta_*}
c_s \, \ud \eta$ is the sound horizon at last scattering. The peak positions
thus depend on the physical densities in matter $\Omega_m h^2$ and
baryons $\Omega_b h^2$ through the sound horizon, and additionally
through curvature and dark energy properties from the angular diameter
distance. The same physics that gives rise to acoustic peaks in the
CMB should produce oscillations in the matter power spectrum. These
baryon oscillations were first detected in early 2005 in the clustering 
of the SDSS luminous red galaxy sample~(Eisenstein et al 2005)
and in the 2dF galaxy redshift sample~(Cole et al 2005).
Observing the connection between the fluctuations
that produced the CMB anisotropies and those responsible for large-scale
structure is an important test of the structure formation paradigm.

The relative heights of the acoustic peaks are influenced by baryon inertia,
gravitational driving and photon diffusion. Baryons reduce the ratio
of pressure to energy density in the plasma, shifting the midpoint
of the oscillations to higher over-densities: $-(1+R)\psi$.
For adiabatic oscillations this enhances the 1st, 3rd etc.\ (compressional)
peaks over the 2nd, 4th etc. 
An additional
effect comes from the gravitational driving term in equation~(\ref{eq:4})
which initially resonantly drives the acoustic oscillations for the short
wavelength modes that underwent oscillation during radiation domination.
Increasing the matter density shifts matter-radiation equality to earlier
times and the resonance is less effective for the low-order acoustic peaks.
In combination, these two effects have allowed accurate measurements
of the matter and baryon densities from the morphology of the
acoustic peaks. From the three-year WMAP data alone,
$\Omega_b h^2 = 0.0223^{+0.0007}_{-0.0009}$ and $\Omega_m h^2 =
0.127^{+0.007}_{-0.01}$ in flat, $\Lambda$CDM
models~(Spergel et al 2006).
The baryon
density implies a baryon-to-photon ratio $(6.10 \pm 0.2)\times 10^{-10}$
and predicts abundances for primordial deuterium, $^3$He and $^4$He
that are consistent with observations.
There is some tension between the low matter density favoured by
WMAP3 and that favoured by tracers of large-scale structure, most
notably weak gravitational lensing~(Spergel et al 2006).
Better measurements of
the third and higher peaks will be very helpful here, and we look
forward to sub-percent level precision in the determination
of densities with the future Planck data\footnote{http://www.rssd.esa.int/index.php?project=Planck}.

With the matter and baryon densities fixed by the morphology of the
acoustic peaks, the acoustic oscillations become a standard ruler with
which to measure $d_A(r_*)$, determined to be
$13.7\pm 0.5$~Gpc~(Spergel et al 2003).
The angular diameter distance depends on the Hubble parameter $H(z)$ and
hence the composition of the universe. The dark energy model, curvature and
sub-eV neutrino masses have no effect on the pre-recombination universe
and they only affect the CMB through $d_A(r_*)$ and their large-scale
clustering through the late-time ISW effect. The discriminatory power
of the latter is limited by cosmic variance leading to the \emph{geometric
degeneracy} between curvature and dark energy~(Efstathiou \& Bond 1999).
For example,
closed models with no dark energy fit the WMAP data but imply a low
Hubble constant and $\Omega_m h$ in conflict with other datasets.
Using the Hubble Space Telescope (HST) measurement of $H_0$, WMAP3 constrains
the spatial sections of the universe to be very close to flat:
$\Omega_K = -0.003^{+0.013}_{-0.017}$ and $\Omega_\Lambda =
0.78^{+0.035}_{-0.058}$ for cosmological constant
models~(Spergel et al 2006).

\vspace{1cm}
\noindent{\bf {4.3 Damping Tail and Photon Diffusion}}\\

The acoustic oscillations in the photon-baryon plasma are exponentially
damped at last scattering on comoving scales $\simless 30$~Mpc due to
photon diffusion~(Silk 1968).
Furthermore, last scattering is not perfectly
sharp: photons last scattered around recombination within a shell of
thickness $\sim 80$~Mpc, and line of sight averaging through this shell
also washes out anisotropy from small-scale fluctuations.
We thus expect an exponential `damping tail' in the temperature
spectrum and this is seen in the ground and balloon-based data in
the left plot in Fig.~\ref{challinor:fig4}.

On scales $l > 2000$, the CBI, operating around 30~GHz,
sees power in excess of that expected
from the primary anisotropies at the $3\sigma$
level~(Bond et al 2005).
An excess is also
seen at smaller angular scales (centred on $l \sim 5000$) with the
BIMA array operating at $28.5$~GHz~(Dawson et al 2006).
Both analyses exclude point-source
contamination as the source of the excess, suggesting instead that
they are seeing a secondary contribution to the anisotropy from
Compton up-scattering of CMB photons off hot gas in unresolved
distant galaxy clusters. This explanation in terms
of the Sunyaev-Zel'dovich (SZ) effect~(Sunyaev \& Zeldovich 1972)
favours a variance in
the matter over-density $\sigma_8 \approx 1$ (with $1\sigma$ errors
at the 20\%\ level), on the high side compared to inferences from current
CMB and large-scale structure data~(Spergel et al 2006).
Optical follow-up of
the BIMA fields shows no (anti-)correlation between galaxy over-densities and
the anisotropy images, but the image statistics are consistent with
SZ simulations. Data from the several high-resolution CMB experiments that
will soon be operational should identify a definitive source for this
excess small-scale power.

\vspace{1cm}
\noindent{\bf {4.4 $E$-mode Polarization and $TE$ Cross-Correlation}}\\

Current measurements of the $E$-mode polarization power spectrum and the
cross-correlation with the temperature anisotropies are fully
consistent with predictions based on the best-fit adiabatic model
to the temperature anisotropies. This is an important test of the structure
formation model: the polarization mainly reflects the
plasma bulk velocities around recombination and these are consistent,
via the continuity equation,
with the density fluctuations that mostly seed the temperature anisotropies.

Apart from constraints on the reionization
optical depth from large-angle polarization data (see Sec.~4.5),
the power of the current polarization data for constraining parameters
in adiabatic, $\Lambda$CDM models is rather limited. 
More important are
the qualitative conclusions that we can draw from the data:
\begin{itemize}
\item The well-defined oscillations in $C_l^{TE}$ further support the phase
coherence of the primordial fluctuations, i.e.\ all modes with a given
wavenumber oscillate in phase. This is a firm prediction of inflation
models since then the fluctuations are produced in the growing mode and
evolve passively, but is at odds with defect models.
\item The (anti-)correlation between the polarization and temperature on degree
scales (see Fig.~\ref{challinor:fig4}) is evidence for fluctuations
at last scattering that are outside the Hubble radius and are
adiabatic~(Peiris et al 2003). This is more direct, model-independent
evidence for such fluctuations than from the temperature anisotropies since
the latter could have been produced on these scales gravitationally all along
the line of sight.
\item The peak positions in polarization, as for
the temperature, are in the correct locations for adiabatic initial
conditions. Pre-WMAP3 analyses, combining CMB temperature and polarization
with large-scale structure and nucleosynthesis priors on the baryon density,
limit the contribution from isocurvature initial
conditions to the CMB power to be less than 30\%, allowing for the
most general correlated initial conditions~(Dunkley et al 2005).
\item That $E$-mode power peaks at the minima of the temperature
power spectrum increases our confidence that the primordial power
spectrum is a smooth function with no features `hiding' on scales
that reach a mid-point of their acoustic oscillation at last scattering (and
so contribute very little to the temperature anisotropies).
\end{itemize}

\vspace{1cm}
\noindent{\bf {4.5 Large-Angle Polarization from Reionization}}\\

The large-angle polarization generated by re-scattering at reionization
was first seen in the $TE$ correlation in the first-year WMAP
data~(Kogut et al 2003).
This provided a broad constraint on the optical depth with mean $\tau=0.17$
and prompted a flurry of theoretical activity to explain such early
reionization. With the full polarization analysis of the three-year
release~(Page et al 2006),
the reionization signal can be seen in the large-angle $C_l^{E}$ spectrum
(see the insert in the bottom right plot in Fig.~\ref{challinor:fig4}).
To obtain this spectrum required aggressive cleaning of Galactic foregrounds,
using the 22.5-GHz (K-band) channel as a template for polarized synchrotron
emission and a model for the contribution of polarized thermal dust
emission, as well as a careful treatment of correlated noise in the
noise-dominated polarization maps. However, the observation that the $B$-mode
spectrum from the same analysis is consistent with zero suggests that
residual foreground contamination is under control. On the basis of the
$EE$ spectrum alone, the WMAP team find $\tau=0.10 \pm 0.03$, considerably
lower than the best-fit to the one-year $TE$ spectrum. The value from
the improved three-year analysis sits much more comfortably with astrophysical
reionization models.

\vspace{1cm}
\noindent{\bf {5. CMB Constraints on Inflation}}\\

Inflation is a posited period of accelerated expansion in the early universe.
It was originally proposed as a solution to a number of now-classic problems
with Friedmann cosmologies, such as why the three-geometry is now so close to
Euclidean and why the CMB temperature is so uniform given that points
at last scattering subtending more than one degree today should never
have been in causal contact~(Guth 1981).
In the inflationary picture,
the observable universe is believed to derive from a small, causally-connected
region that was inflated by at least 60 e-folds during a phase of
accelerated expansion. One mechanism to realise inflation, which
requires a violation of the weak energy condition, is with
a scalar field $\phi$ --- the inflaton --- evolving slowly over a flat
part of its interaction potential $V(\phi)$. It has proved difficult to
realise inflation from (fundamental) field theory, and this has led to a
plethora of phenomenological models in the literature. String-inspired
approaches, in which the inflaton may emerge as one of the moduli fields
in the low-energy effective theory are also being actively pursued,
e.g.\ Kachru et al (2003).

\vspace{1cm}
\noindent{\bf {5.1 Inflationary Power Spectra}}\\

Inflation naturally predicts a universe that is very close to flat, consistent
with the observed positions of the CMB acoustic peaks as discussed
in Sec.~4. Significantly, it also naturally provides a causal mechanism for
generating initial curvature fluctuations~(Bardeen, Steinhardt \& Turner 1983)
and gravitational waves~(Starobinskii 1979)
with almost scale-invariant, power-law spectra.
The mechanism is an application of semi-classical quantum gravity, in which
the perturbations in the inflaton field and metric fluctuations are
quantised on a classical background that is close to de Sitter. Since
the physical wavelength of a mode gets pushed outside the Hubble radius by
the accelerated expansion, vacuum fluctuations initially deep inside the
Hubble radius are stretched to cosmological scales and amplified during
inflation. The spectra of the curvature perturbations and gravity waves
can be approximated as power laws over the range of scales relevant for
cosmology:
\begin{equation}
\mathcal{P}_{\mathcal{R}} \approx A_s (k/k_0)^{n_s-1} \qquad , \qquad
\mathcal{P}_{h} \approx A_t (k/k_0)^{n_t} \, ,
\label{eq:6}
\end{equation}
where the amplitudes and spectral indices are given by [see e.g.\
Lidsey et al (1997) and references therein]
\begin{equation}
A_s = \frac{H^2}{\pi \epsilon m_{\mathrm{Pl}}^2}\, , \quad
n_s - 1 = -4\epsilon + 2\eta \, , \quad
A_t \equiv r A_s = \frac{16 H^2}{\pi m_{\mathrm{Pl}}^2} \, , \quad
n_t = - 2 \epsilon \, .
\label{eq:7}
\end{equation}
Here, $H$ is the Hubble parameter during inflation evaluated when
the mode $k_0$ exits the Hubble radius, i.e.\ when $k_0=aH$, and
$\epsilon$ and $\eta$ are (Hubble) slow-roll parameters that
depend on the evolution of $H$ through inflation.
The potential
energy dominates the stress-energy of the scalar field during inflation,
and in this limit $H$, $\epsilon$ and $\eta$ are related to the potential
$V(\phi)$ by
\begin{equation}
H^2 \approx \frac{8\pi}{3m_{\mathrm{Pl}}^2}V \, , \quad
\epsilon \approx \frac{m_{\mathrm{Pl}}^2}{16\pi} \left(\frac{V'}{V}\right)^2
\, , \quad
\eta \approx \frac{m_{\mathrm{Pl}}^2}{8\pi}\left[ \frac{V''}{V}-
\frac{1}{2} \left(\frac{V'}{V}\right)^2\right] \, ,
\label{eq:8}
\end{equation}
where primes denote derivatives with respect to the field $\phi$.
Equations~(\ref{eq:7}) and (\ref{eq:8}) imply that the power spectrum of
gravitational waves from slow-roll inflation depends only the energy
density $V$. This is often expressed in terms of an \emph{energy scale of
inflation} $E_{\mathrm{inf}} = V^{1/4}$, which gives a tensor-to-scalar
ratio
\begin{equation}
r = 8\times 10^{-3} (E_{\mathrm{inf}}/10^{16}\, \mathrm{GeV})^4
\label{eq:9}
\end{equation}
for a scalar amplitude $A_s = 2.36 \times 10^{-9}$. The four observables,
$A_s$, $A_t$, $n_s$ and $n_t$ are not independent since they derive
from three parameters, $H$, $\epsilon$ and $\eta$. This results in the
leading-order slow-roll consistency relation $r=-8n_t$. Verifying
this observationally would be a remarkable triumph for inflation, but
the prospects are poor even after accounting for the long lever-arm that a
combination of CMB and direct detections could
provide~(Smith, Peiris \& Cooray 2006).

\begin{figure}
\begin{center}
\includegraphics[width=0.7\textwidth,angle=0]{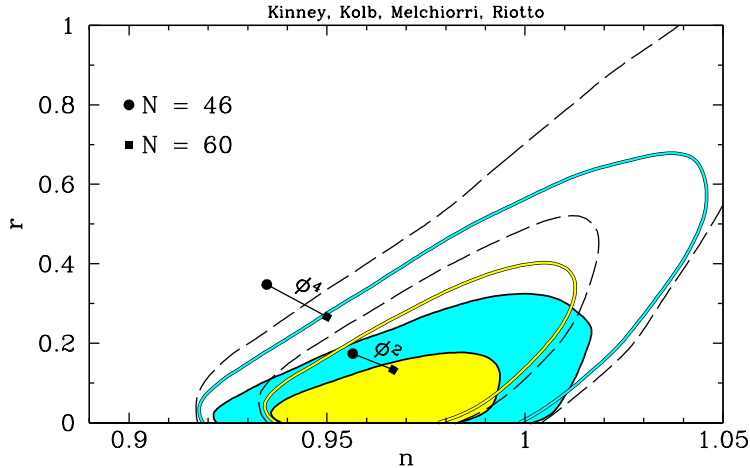}
\end{center}
\caption{\label{challinor:fig5} Constraints in the $r$-$n_s$ plane for
models with no running from Kinney et al (2006).
Blue contours are
68\%\ confidence regions and yellow are 95\%. The filled contours are
from combining WMAP3 and the SDSS galaxy survey; open are with WMAP3 alone.
These results assume the HST prior on $H_0$; dropping this prior gives
the dashed contours. The predictions for $V \propto \phi^2$ and $\phi^4$ are
shown assuming that modes with $k=0.002\,\mathrm{Mpc}^{-1}$ left the Hubble
radius between 46 and 60 e-folds before the end of inflation.
}
\end{figure}

Constraints in the $r$-$n_s$ plane from WMAP3 and the SDSS galaxy survey are
shown in Fig.~\ref{challinor:fig5}, taken from Kinney et al (2006).
The point $r=0$ and $n_s=1$ corresponds to inflation occurring at low
energy with essentially no evolution in $H$ (and hence a very flat potential);
the gravitational waves are negligible and the curvature fluctuations
have no preferred scale. This \emph{Harrison-Zel'dovich} spectrum
is clearly disfavoured by the data, but is not yet excluded at the
95\%\ level. Attempts to pin down $n_s$ with current CMB temperature data are
still hampered by a degeneracy between $n_s$, $A_s$, the reionization optical
depth $\tau$ and the baryon density~(Lewis 2006).
The WMAP3 measurement of
$\tau$ from large-angle polarization helps considerably in breaking this
degeneracy,
and leads to a marginalised constraint of $n_s = 0.987^{+0.019}_{-0.037}$
in inflation-inspired models~(Spergel et al 2006).
The 95\%\ upper limit on
the tensor-to-scalar ratio from WMAP3 and SDDS is $0.28$ for power-law spectra,
thus limiting the inflationary energy scale
$E_\mathrm{inf} < 2.4\times 10^{16}\,\mathrm{GeV}$.
We see from Fig.~\ref{challinor:fig5} that large-field models with
monomial potentials $V(\phi) \propto \phi^p$ are now excluded at high
significance for $p\geq 4$.

Slow-roll inflation
predicts that any running of the spectral indices with scale should
be second-order in the slow-roll parameters, i.e.\ $O[(n_s-1)^2]$.
The CMB alone provides a rather limited lever-arm for measuring
running, with current data having very little constraining power for
$k > 0.05\,\mathrm{Mpc}^{-1}$ (corresponding to $l\sim 700$). However,
there is persistent, though not yet compelling, evidence for running from the
CMB:
WMAP3 alone gives
$\ud n_s / \ud \mathrm{ln} k = -0.102^{+0.05}_{-0.043}$~(Spergel et al 2006),
allowing for gravitational waves.
Running near this mean value would be problematic for slow-roll
inflation models. The tendency for the CMB to favour large negative running
is driven by the large-angle ($l \simless 15$) temperature data. A more
definitive assessment of running must await independent
verification of the large-scale spectrum from Planck and improved small-scale
data from a combination of Planck and further ground-based observations.
The current evidence for running weakens considerably when small-scale
data from the Lyman-$\alpha$ forest (i.e.\ absorption lines in quasar spectra
due to neutral hydrogen in the intergalactic medium) is
included~(Seljak, Slosar \& McDonald 2006; Viel, Haehnelt \& Lewis 2006).
The Ly-$\alpha$ data probes the quasi-linear fluctuations on
$\sim \mathrm{Mpc}$ scales at redshifts around 3, and is sensitive to the
amplitude and slope of the matter power spectrum at these redshifts and
scales. There is currently some tension between the amplitudes inferred from
Ly-$\alpha$ and CMB, and this is worsened by inclusion of the negative running
favoured by the CMB.

\vspace{1cm}
\noindent{\bf {5.3 Non-Gaussianity and Inflation}}\\

There are further predictions of slow-roll inflation that are amenable
to observational tests. The fluctuations from single-field models
should be adiabatic and any departures from Gaussian statistics should
be unobservably small. Gaussianity follows, in part, from the requirement
of a flat potential and hence small self-interactions if inflation
is to happen [see Bartolo et al (2004) for a recent review].
However, adiabaticity and Gaussianity can be violated in models
with several scalar fields.
An example of the latter is the curvaton model~(Lyth \& Wands 2002),
which can produce a large correlated isocurvature mode and observably-large
non-Gaussianity if the curvaton field decays before its energy density
dominates that of radiation. As we have already noted, current
data do allow a sizeable isocurvature fraction, but this is not
favoured. In many models, the non-Gaussian curvature perturbation
can be written as the sum of a Gaussian part plus the square of a Gaussian:
symbolically
\begin{equation}
\mathcal{R} = \mathcal{R}_{\mathrm{G}} + f_{\mathrm{NL}}^{\mathcal{R}}
(\mathcal{R}_{\mathrm{G}}^2 - \langle \mathcal{R}_{\mathrm{G}}^2 \rangle) ,
\end{equation}
where, in general, $f_{\mathrm{NL}}^{\mathcal{R}}$ is scale-dependent and
the quadratic part is a convolution. Single-field slow-roll inflation
predicts
$f_{\mathrm{NL}}^{\mathcal{R}} \sim O(\epsilon)$~(Maldacena 2003),
but it can be much higher in alternative models. Observational
constraints are usually expressed in terms of the $f_{\mathrm{NL}}$
appropriate to the gravitational potential $\psi$ at last scattering.
For $f_{\mathrm{NL}}^{\mathcal{R}} \gg 1$, we have
$f_{\mathrm{NL}} \approx -5 f_{\mathrm{NL}}^{\mathcal{R}}/3$ since
$O(1)$ non-linear corrections in the relation of the curvature
to metric perturbation can then be ignored. The best constraints
on a scale-independent $f_{\mathrm{NL}}$ are from an analysis of the
three-point function of the three-year WMAP maps:
$-54 < f_{\mathrm{NL}} < 114$ at 95\%\ confidence~(Spergel et al 2006).
Planck data should have sensitivity down to
$f_{\mathrm{NL}} \sim 5$~(Komatsu \& Spergel 2001),
but this is still too large to
expect to see anything in simple inflation models.

\vspace{1cm}
\noindent{\bf {6. Searching for Gravity Waves with the CMB}}\\

A detection of a Gaussian-distributed background of gravitational waves with
cosmological wavelengths and a nearly scale-invariant (but red)
spectrum would be seen
by many as compelling evidence that inflation occurred. The amplitude
of this background is directly related to the energy scale of inflation,
and the near scale-invariance follows from the slow decrease in the Hubble
parameter during inflation. Of course, a non-detection would not rule out
inflation having happened at a low enough energy, but, importantly, a
detection would
rule out some alternative theories for the generation of the curvature
perturbation, such as the cyclic model~(Steinhardt \& Turok 2002),
that predict negligible
gravity waves. The large-angle $B$ mode of CMB polarization is a promising
observable with which to search for the imprint of gravity waves
since a detection would not be confused by linear curvature perturbations.

%
\begin{figure}
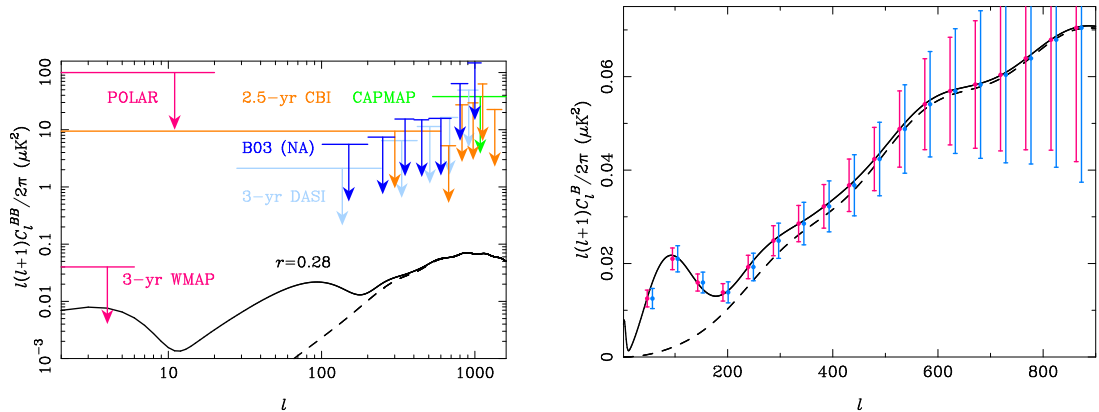

\begin{center}
\includegraphics[width=0.36\textwidth,angle=-90]{current_data_B_0306.ps}
\includegraphics[width=0.36\textwidth,angle=-90]{Clover_errors_Chile2yr_4x8p9deg_rp28_B.ps}
\end{center}
\caption{\label{challinor:fig6} Left: current 95\%\ upper limits on the
$B$-mode polarization power spectrum. The solid line is the
theoretical prediction for a tensor-to-scalar ratio $r=0.28$ --- the current
95\%\ limit from the temperature power spectrum and 
galaxy clustering~(Spergel et al 2006) --- while
the dotted line is the contribution from weak gravitational lensing.
Right: error forecasts for Clover after a two-year campaign observing
1000~deg$^2$ divided between four equal-area fields. Blue error bars
properly account for $E$-$B$ mixing effects due to the finite sky coverage
while magenta ignore this. The tensor-to-scalar ratio is again $r=0.28$.}
\end{figure}

In Fig.~\ref{challinor:fig6} we show a compilation of current direct upper
limits on the $B$-mode power spectrum. The solid curve is the theoretical
spectrum, including the contribution from weak gravitational lensing, for
$r=0.28$ --- the 95\%\ upper limit inferred from WMAP3 temperature and
$E$-mode data and SDSS galaxy clustering~(Spergel et al 2006).
Clearly, the
direct measurements are not yet competitive, with at least a factor ten
improvement in sensitivity required.
The $2\sigma$ limit to
determining $r$ from ideal CMB temperature observations is $0.14$ and this
improves to $0.04$ with $E$-mode data. Although the WMAP temperature data is
already cosmic-variance limited on scales where gravity waves contribute, the
constraints on $r$ are considerably worse than $r=0.14$ due to the uncertainties
in other cosmological parameters. In principle, $B$-mode measurements of
$r$ can do much better as they are limited only by how well the lensing
signal can be subtracted. Lensing reconstruction methods based on the
non-Gaussian action of lensing on the CMB have been
proposed, e.g.\ Hu (2001), and
with the most optimal methods $r\sim 10^{-6}$ may be 
achievable~(Seljak \& Hirata 2003).
In practice, astrophysical foregrounds and instrumental systematic effects
are likely to be a more significant obstacle.

A number of groups are now designing and constructing a new generation
of CMB polarimeters that aim to be sensitive down to
$r\sim 0.01$.
These should be reporting data within the next
five years, a timescale similar to the Planck satellite.
The constraint $r<0.28$ gives an r.m.s.\ gravity wave contribution
$< 200\, \mathrm{nK}$ to $B$-mode polarization. Detecting such signals
requires instruments with many hundreds,
or even thousands, of detectors, and demands
exquisite control of instrumental effects and broad frequency coverage
to deal with polarized Galactic foreground emission. With the exception
of SPIDER, which will aim to survey around half of the sky,
the surveys will each target small sky areas ($\simless 1000\,\mathrm{deg}^2$)
in regions of low foreground emission in total intensity. Despite this,
removing foregrounds to the 10\%\ level will likely be required to see
a $B$-mode signal at $r=0.01$. As an example of a next-generation
instrument, we show in Fig.~\ref{challinor:fig6} projected errors
on the $B$-mode power spectrum from the UK-led Clover experiment.
Clover will have over 1200 super-conducting detectors distributed over
three scaled telescopes centred on 97, 145 and 225~GHz, each with
better than 10-arcmin resolution.
The instrument is planned to be deployed
at the Chajnantor Observatory, Chile. The error forecasts in
Fig.~\ref{challinor:fig6} are for a tensor-to-scalar ratio $r=0.28$ for
direct comparison with the adjacent plot of current upper limits. However,
the Clover survey is optimised for smaller $r\sim 0.01$, and is designed
to be limited on large scales by sample variance of the lens-induced $B$
modes after two years of operation.

\vspace{1cm}
\noindent{\bf {7. Other Physics in the CMB Fluctuations}}\\

Within the $\Lambda$CDM model, the major remaining CMB milestones are the
detection of gravitational secondary effects (weak lensing and the
non-linear ISW effect from collapsing structures), various scattering secondary
effects from bulk velocities around and after the epoch of reionization,
and the detection of $B$-mode polarization and (possibly) gravitational
waves. A number of deep surveys at arcminute resolution will soon commence
to study the temperature anisotropies at high $l$. Their main goal
is to characterise the scattering secondaries, and hence learn more about
the reionization history and morphology, and to detect the gravitational
lensing effect in the temperature anisotropies. At the same time,
high-sensitivity polarization surveys are being undertaken and these
should also provide further valuable information on the weak-lensing effect.
From the viewpoint of fundamental physics, the main interest in such
small-scale observations is the possibility of using CMB lensing to determine
neutrino
masses and further constrain the dark-energy model. However, we should
also be mindful of the possibility of serendipitous discovery of other
physics in
the small-scale CMB fields, such as the imprint of cosmic strings or primordial
magnetic fields. We now discuss some of these briefly.

Cosmology has the potential to place constraints on the absolute
neutrinos masses
rather than the (squared) differences from neutrino oscillations
[see Lesgourgues \& Pastor (2006) for a recent review].
The current constraint on the sum of neutrino masses from CMB, galaxy
clustering and Lyman-$\alpha$ forest data is $\sum m_\nu < 0.17\,\mathrm{eV}$
at 95\%\ confidence~(Seljak et al 2006).
The implied sub-eV masses mean neutrinos
are relativistic at recombination and their effect on the CMB is limited to
late times. Changes they induce in the angular diameter distance are degenerate
with dark energy and so, to constrain masses from the CMB alone, we need
to look to gravitational lensing. Non-relativistic massive neutrinos
increase the expansion rate over massless one impeding the growth of structure.
This effect is cancelled on scales larger than the neutrino comoving Jeans'
length (which decreases in time and is inversely proportional to the
mass) by neutrino clustering, but on smaller scales the growth of
fluctuations in the matter density is slowed. It is this
suppression that allows the energy density (or sum of masses) of neutrinos
to be constrained by small-scale tracers of matter clustering.
The effect of neutrino masses on the power spectrum of the lensing deflections
is scale dependent: no effect at low $l$ but a reduction in power at high
$l$.
To exploit the effects of neutrino physics in
CMB lensing, one must first attempt to reconstruct the underlying lensing
deflection field from the lensed CMB fields. An accurate
reconstruction requires high resolution and is greatly helped by polarization
measurements once the sensitivity is high enough to image lens-induced
$B$ modes~(Hu \& Okamoto 2001; Seljak \& Hirata 2003).
Assuming a normal hierarchy of neutrino masses with
two essentially massless\footnote{This combination has the smallest
possible neutrino energy density and hence cosmological effect.},
Kaplinghat, Knox \& Song (2003)
estimated that the mass of the third
should be measurable to an accuracy of $0.04\,\mathrm{eV}$ with a future
polarization satellite mission. Errors of this magnitude are comparable
to what should be achievable in the future with galaxy lensing, but
with quite different potential systematic effects.
It is an interesting result since atmospheric
neutrino oscillations then imply that a detection of mass with the CMB
\emph{must} be possible at the $1\sigma$ level. Of course, the significance
will be higher if the lightest neutrinos are not massless, or in the inverted
hierarchy. Note that the latter is on the verge of being ruled out with
the current cosmological constraints~(Seljak et al 2006).
It is also interesting to
question whether CMB lensing will be able to tell us anything about
individual masses rather than their sum? Taking differences from oscillation
data at face value, the answer appears to be that dropping the assumption
of degenerate masses would produce an improved fit to an idealised,
cosmic-variance-limited reconstruction of the lensing power spectrum if
$\sum m_\nu < 0.1 \, \mathrm{eV}$~(Slosar 2006).
However, attempting to measure
the mass differences with no prior from atmospheric oscillations is
not possible because of degeneracies with other parameters and these severely
degrade ones ability to measure $\sum m_\nu$ without mass priors.

The details of the dark-energy sector affect the primary CMB anisotropies
only through the angular diameter distance and the late-time ISW effect.
Using the former is plagued by degeneracies while the latter is hampered
by cosmic variance. The effect of dark energy on CMB lensing is felt almost
exclusively through the change in the expansion rate which is independent
of scale. The different scale dependences from dark energy (say
to changes in the equation of state parameter $w=p/\rho$)
and massive neutrinos should allow them to be measured separately with the
lensed CMB.
Kaplinghat et al (2003) find a marginalised $1\sigma$ error on $w$ of
$0.18$ from a future polarization satellite. Of course, tomography proper
is not possible with the fixed source plane of the CMB (i.e.\ last scattering),
and the CMB constraints on dark energy will not be competitive with future
galaxy lensing and clustering (via baryon oscillations) surveys.

A significant feature of recent attempts to realise inflation in
string/M-theory cosmology is the recognition that (local) cosmic strings
may be produced generically at the end of brane
inflation~(Sarangi \& Tye 2002).
The details of the strings network (such as the spectrum of
tensions and inter-commutation rates) depend on the details of the brane
scenario, but the tensions $G\mu / c^4$ plausibly exceed $10^{-11}$.
Cosmic strings leave an imprint in the CMB temperature anisotropies
due to string wakes stirring up the plasma prior to recombination, and
from the integrated effect of rapidly moving strings crossing the line
of sight~(Kaiser \& Stebbins 1984).
Current data limits the contribution of local strings
to the temperature power spectrum to be $\simless 10\%$, corresponding
to $G\mu/c^4 < 2.7\times 10^{-7}$~(Seljak et al 2006).
Future high-resolution temperature data
should improve this bound on the tension further, and searches
for stringy non-Gaussian imprints in CMB maps should also help.
As with the search for inflationary gravitational waves, $B$-mode polarization
may prove to be the most promising observable for constraining strings.
String networks excite scalar, vector and tensor perturbations and the
latter two lead to $B$-mode polarization from the epoch of recombination
and reionization. Many brane-inflation models predict a negligible gravity
wave production \emph{during} inflation in which case strings should
be the dominant primordial source of $B$-mode polarization.
Seljak \& Slosar (2006)
argue that $B$-mode measurements with a future polarization satellite
may improve on current string constraints by an order of magnitude.
The string signal peaks around $l\sim 1000$ and so the hope is that
observations covering a range of scales should be able to separate it from
primordial gravity waves and the lensing signal. However, further work is
required to extend and test lensing reconstruction methods in the presence of
a possible non-Gaussian string signal.

Finally, we note that there are statistically-significant anomalies
in the large-angle temperature anisotropies, as imaged by
COBE~(Smoot 1992) and WMAP~(Hinshaw 2003), that may signal 
departures from rotational invariance and/or Gaussianity; for
a recent review and WMAP3 analysis, see Copi et al (2006)
and references therein.
Arguably most significantly, the $l \leq 6$ multipoles seem to favour a
preferred axis about which they maximise the power concentrated in a single
$m$ mode~(Land \& Magueijo 2005).
There are also significant correlations of the
quadrupole and octupole ($l=3$) with the ecliptic plane and the direction
of the equinoxes and/or CMB dipole~(Copi et al 2006).
The significance of these
anomalies is still under debate, as is their possible explanation. Suggestions
include unidentified instrumental effects, residual foreground contamination,
and effects of the local universe~(Vale 2005),
although it appears unlikely that the last
two can be responsible~(Cooray \& Seto 2005).
There have also been a number of suggestions that
the large-angle anomalies may have a more fundamental origin, such as
a topologically small universe~(de Oliveira-Costa et al 2004)
or non-fluid dark energy~(Battye \& Moss 2006).
Independent
verification with the Planck data and improved analyses of further years
of WMAP data should help with tracking down the source of these large-angle
effects\footnote{The first-year WMAP data was already signal-dominated on
large angular scales so further integration helps not by improving the
signal-to-noise but by bettering our understanding of instrumental and foreground
effects.}.

\vspace{1cm}
\noindent{\bf {8. Summary}}\\

Many of the bold predictions of CMB physics have now been impressively
verified with a large number of independent observations. The large-scale
Sachs-Wolfe effect, acoustic peak structure, damping tail, late-time
integrated Sachs-Wolfe effect, $E$-mode polarization and the effect
of reionization have all been detected. The large-scale anisotropies and
the first three acoustic peaks have now been measured accurately and have
yielded impressive constraints on cosmological parameters. The data is
consistent with a very simple cosmological model with adiabatic
primordial fluctuations, with an almost scale-free spectrum, evolving
passively in a spatially-flat, $\Lambda$CDM universe.

Inflation continues to stand up to exacting comparisons with both CMB
and tracers of matter clustering. Evidence for dynamics during inflation
is emerging, most notably from the recent third-year WMAP data: models with
scale-invariant curvature perturbations and no gravity waves are
on the brink of being ruled out at 95\%\ confidence. There are hints of
a run in the spectral index index in current CMB data at a level that
would be problematic for many inflation models, but this is not corroborated
by probes of the matter power spectrum on small scales (the Lyman-$\alpha$
forest). 
In the near future
we can expect better measurements of the third acoustic peak in the
temperature anisotropies and beyond. With Planck we can expect
a per-cent level determination of the spectral index of curvature perturbations
and a much more definitive
assessment of running and any potential conflict with the small-scale
matter power spectrum.

We look forward to improvements in $E$-mode polarization data and the
better constraints this will bring on non-standard cosmological models such
as those with a significant contribution from isocurvature fluctuations.
On a similar timescale, a new generation of small-scale temperature
experiments should constrain further the reionization history and its
morphology, and detect the effect of weak gravitational lensing by
large-scale structure in the CMB. Looking a little further ahead,
a new generation of high-sensitivity polarization-capable instruments
have the ambition of detecting the imprint of gravitational waves from
inflation. They should be sensitive down to tensor-to-scalar ratios
$r\sim 0.01$ --- corresponding to an energy scale of inflation around
$1.0 \times 10^{16} \,\mathrm{GeV}$ --- and
will place tight constraints on inflation
models. There is also exciting secondary science that can
be done with these instruments, such as lensing reconstruction which
brings with it the promise of competitive constraints on neutrino masses
from the CMB alone. Finally, there is always the hope of serendipitous
discovery, such as the non-Gaussian signature of cosmic strings,
perhaps produced at the end of brane inflation, in small-scale temperature
maps. These continue to be exciting times for CMB research.

\vspace{1cm}
\noindent{\bf Acknowledgements} \\

AC thanks the Royal Society for a University Research Fellowship,
the organisers for the invitation to attend this stimulating workshop
and the British Council for sponsoring my attendance. Thanks also to
Will Kinney and Bill Jones for permission to include their figures.

\vspace{1cm}
\noindent {\bf {References}}\\

\rfnce Bardeen, J.M., Steinhardt, P.J., Turner, M.S., 1983, Phys.\ Rev.\ D,
28, 679

\rfnce Battye, R.A., Moss, A., 2006, arXiv:astro-ph/0602377

\rfnce Bartolo, N.\ et al, 2004, Phys.\ Rept., 402, 103

\rfnce Bennett, C.L.\ et al, 2003, ApJS, 148, 1

\rfnce Bond, J.R.\ et al, 2005, ApJ, 626, 12

\rfnce Bough, S., Crittenden, R., 2004, Nature, 427, 45

\rfnce Cabre, A.\ et al, 2006, arXiv:astro-ph/0603690

\rfnce Challinor, A., 2005, in {\it The Physics of the Early Universe},
E.\ Papantonopoulos (ed.), Lect.\ Notes.\ Phys. 653, 71

\rfnce Cole, S.\ et al, 2005, MNRAS, 362, 505

\rfnce Cooray, A., Seto, N., 2005, JCAP, 12, 4

\rfnce Copi, C.\ et al, 2006, arXiv:astro-ph/0605135

\rfnce Dawson, K.S.\ et al, 2006, arXiv:astro-ph/0602413

\rfnce Dunkley, J.\ et al, 2005, Phys.\ Rev.\ Lett., 95, 261303

\rfnce Eisenstein, D.J.\ et al, 2005, ApJ, 633, 560

\rfnce Efstathiou, G., Bond, J.R., 1999, MNRAS, 304, 75

\rfnce Guth, A.H., 1981, Phys.\ Rev.\ D, 23, 347

\rfnce Hinshaw, G.\ et al, 1996, ApJL, 464, 17

\rfnce Hinshaw, G.\ et al, 2006, arXiv:astro-ph/0603451

\rfnce Hu, W., 2001, ApJL, 557, 79

\rfnce Hu, W., 2002, Ann.\ Phys., 303, 203

\rfnce Hu, W., Sugiyama, N., 1995, ApJ, 444, 489

\rfnce Hu, W., White, M., 1997, Phys.\ Rev.\ D, 56, 596

\rfnce Hu, W., Dodelson, S., 2002, Ann.\ Rev.\ Astron.\ Astrophys., 40, 171

\rfnce Hu, W., Okamoto, T., 2002, ApJ, 574, 566

\rfnce Jones, W.C.\ et al, 2005, arXiv:astro-ph/0507494

\rfnce Kachru, S.\ et al, 2003, JCAP, 0310, 013

\rfnce Kaiser, N., Stebbins, A., 1984, Nature, 310, 391

\rfnce Kamionkowski, M., Kosowsky, A., Stebbins, A., 1997,
Phys.\ Rev.\ D, 55, 7368

\rfnce Kaplinghat, M., Knox, L., Song, Y.S., 2003, Phys.\ Rev.\ Lett.,
91, 241301

\rfnce Kinney, W.H.\ et al, 2006, arXiv:astro-ph/0605338

\rfnce Kogut, A.\ et al, 2003, ApJS, 148, 161

\rfnce Komatsu, E., Spergel, D.N., 2001, Phys.\ Rev.\ D, 63, 063002

\rfnce Land, K., Magueijo, J., 2005, Phys.\ Rev.\ Lett., 95, 071301

\rfnce Lesgourgues, J., Pastor, S., 2006, arXiv:astro-ph/0603494

\rfnce Lewis, A., 2006, arXiv:astro-ph/0603753

\rfnce Lewis, A., Challinor, A., 2006, Phys.\ Rept., 429, 1

\rfnce Lidsey, J.E.\ et al, 1997, Rev.\ Mod.\ Phys., 69, 373

\rfnce Lyth, D.H., Wands, D., 2002, Phys.\ Lett.\ B, 524, 5

\rfnce Maldacena, J., 2003, J.\ High Energy Phys., 5, 13

\rfnce Mather, J.C.\ et al, 1994, ApJ, 420, 439

\rfnce de Oliveira-Costa, A.\ et al, 2004, Phys.\ Rev.\ D, 69, 063516

\rfnce Page, L.\ et al, 2006, arXiv:astro-ph/0603450

\rfnce Peiris, H.V.\ et al, 2003, ApJS, 148, 213

\rfnce Rees, M.J., 1968, ApJL, 153, 1

\rfnce Sachs, R., Wolfe, A., 1967, ApJ, 147, 735

\rfnce Sarangi, S., Tye, S.H.H., 2002, Phys.\ Lett.\ B, 536, 185

\rfnce Seljak, U., Hirata, C.M., 2004, Phys.\ Rev.\ D, 69, 043005

\rfnce Seljak, U., Slosar, A., 2006, arXiv:astro-ph/0604143

\rfnce Seljak, U., Slosar, A., McDonald, P., 2006, arXiv:astro-ph/0604335

\rfnce Silk, J., 1968, ApJ, 151, 459

\rfnce Slosar, A., 2006, arXiv:astro-ph/0602133

\rfnce Smith, T.L., Peiris, H.V., Cooray, A., 2006, arXiv:astro-ph/0602137

\rfnce Smoot, G.F.\ et al, 1992, ApJL, 396, 1

\rfnce Spergel, D.N.\ et al, 2003, ApJS, 148, 175

\rfnce Spergel, D.N.\ et al, 2006, arXiv:astro-ph/0603449

\rfnce Starobinskii, A.A., 1979, JETP Lett., 30, 682

\rfnce Steinhardt, P.J., Turok, N., 2002, Science, 296, 1436

\rfnce Sunyaev, R.A., Zeldovich, Y.B., 1972,
Comm.\ Astrophys.\ Space Phys., 4, 173

\rfnce Vale, C., 2005, arXiv:astro-ph/0509039

\rfnce Viel, M., Haehnelt, M.G., Lewis, A., 2006, arXiv:astro-ph/0604310

\rfnce Zaldarriaga, M., 1997, Phys.\ Rev.\ D, 55, 1822

\rfnce Zaldarriaga, M., Seljak U., 1996, Phys.\ Rev.\ D, 55, 1830

\rfnce Zaldarraiaga, M., Seljak, U., 1998, Phys.\ Rev.\ D, 58, 023003

\end{document}